\documentclass[pdflatex,sn-aps]{sn-jnl}

\usepackage{graphicx}%
\usepackage{multirow}%
\usepackage{amsmath,amssymb,amsfonts}%
\usepackage{amsthm}%
\usepackage{mathrsfs}%
\usepackage[title]{appendix}%
\usepackage{xcolor}%
\usepackage{textcomp}%
\usepackage{manyfoot}%
\usepackage{booktabs}%
\usepackage{algorithm}%
\usepackage{algorithmicx}%
\usepackage{algpseudocode}%
\usepackage{listings}%

\theoremstyle{thmstyleone}%
\theoremstyle{thmstyletwo}%

\theoremstyle{thmstylethree}%

\begin{document}

\title[Article Title]{Towards Less Biased Data-driven Scoring with Deep Learning-Based End-to-end Database Search in Tandem Mass Spectrometry}


\author[1]{\fnm{Yonghan} \sur{Yu}}\email{y443yu@uwaterloo.ca}

\author*[1]{\fnm{Ming} \sur{Li}}\email{mli@uwaterloo.ca}

\affil*[1]{\orgdiv{Cheriton School of Computer Science}, \orgname{University of Waterloo}, \orgaddress{\street{200 University Ave W}, \city{Waterloo}, \postcode{N2L3G1}, \state{Ontario}, \country{Canada}}}




\abstract{Peptide identification in mass spectrometry-based proteomics is crucial for understanding protein function and dynamics. Traditional database search methods, though widely used, rely on heuristic scoring functions and statistical estimations have to be introduced for a higher identification rate. Here, we introduce DeepSearch, the first deep learning-based end-to-end database search method for tandem mass spectrometry. DeepSearch leverages a modified transformer-based encoder-decoder architecture under the contrastive learning framework. Unlike conventional methods that rely on ion-to-ion matching, DeepSearch adopts a data-driven approach to score peptide spectrum matches. DeepSearch is also the first deep learning-based method that can profile variable post-translational modifications in a zero-shot manner. We showed that DeepSearch's scoring scheme expressed less bias and did not require any statistical estimation. We validated DeepSearch's accuracy and robustness across various datasets, including those from species with diverse protein compositions and a modification-enriched dataset. DeepSearch sheds new light on database search methods in tandem mass spectrometry.}





\maketitle

\section{Main}\label{sec1}
Peptide identification in mass spectrometry-based (MS-based) proteomics is a fundamental challenge in proteomics~\cite{Nesvizhskii2010-bc}. In one widely adopted MS-based proteomics, proteins are digested into peptides with protease, and the resulting peptides are analyzed by liquid chromatography coupled with tandem MS (MS/MS)~\cite{Aebersold2003-ip}. MS/MS spectra contain mass and intensity information of measured peptide fragment ions. The most widely adopted method for peptide identification is database search, which matches the experimental MS/MS spectra to theoretical spectra deduced from a peptide sequence database ~\cite{Eng2011-nc}. However, almost all existing database search engines~\cite{MaxQuant-kc,MSFragger-fd,msgfplus, xtandem-ze, comet-vg} rely on heuristic scoring functions, most of which match sets of most common ions while ignoring the vast unknown and satellite fragmentations~\cite{Liu2020-xf}. Besides, probabilistic models based on statistical significance~\cite{statical-db, xtandem-ze, MSFragger-fd} or Bayesian probability esitmation~\cite{em-len-statical-db, MaxQuant-kc} have to be introduced to mitigate potential biases of scoring functions for a higher identification rate.

Advancements of deep learning in proteomics boost the amino acid level accuracy of \textit{de novo} peptide sequencing,
which directly infer peptide sequence from MS/MS spectra without any prior information~\cite{DeepNovo-uy,PointNovo, pmlr-casanovo,Mao2023}. DeepNovo~\cite{DeepNovo-uy} introduced spectrum CNN coupled with LSTM to predict peptide sequences from MS/MS spectra. PointNovo~\cite{PointNovo} further improved the accuracy of predicted peptides and enabled resolution-free spectrum encoding with PointNet~\cite{qi2017pointnet}. Previous studies also demonstrated the robustness of transformer-based encoder-decoder architecture by training models across the modalities of MS/MS spectra and peptide sequences~\cite{pmlr-casanovo, Mao2023, prosit_trans}. Nevertheless, most of the existing \textit{de novo} sequencing methods show significant performance drops on datasets with much different protein compositions~\cite{Mao2023}. These methods also fall short of much lower peptide level accuracy and are unable to identify variable post-translational modifications (PTMs), which are crucial in functional and structural analysis of proteins~\cite{Ramazi2021-fy}.

Recently introduced multimodal foundation models under the contrastive learning framework significantly improved the performance in various downstream cross-modal understanding tasks, especially in computer vision and natural language processing~\cite{clip,li2022blip,yu2022coca, onetransformer}. These models are capable of learning a joint embedding space across different modalities and demonstrated profound results in zero-shot learning tasks~\cite{clip, yu2022coca}. Most importantly, the weak supervision regime under these frameworks requires no annotations beyond cross-modal data pairs, demonstrating increased tolerance to biases and enhanced robustness across datasets~\cite{ALIGN, wang2022simvlm}.

Here, we proposed the first end-to-end deep learning-based database search method, DeepSearch. Instead of ion-to-ion matching, DeepSearch employed the cross-modal cosine similarity as the scoring scheme. DeepSearch was trained under the contrastive learning framework and jointly optimized with a \textit{de novo} sequencing objective on MassIVE v2~\cite{MassIVE}, a high-quality set of peptide-spectrum matches (PSMs) built upon human MS/MS library. We demonstrated that the scores reported by DeepSearch are less biased, and our method accurately identified peptides across multiple MS/MS datasets from species with diverse protein compositions. Finally, we highlighted DeepSearch's capacity to report PTM profiles of high accuracy on a phosphorylation-enriched HeLa dataset in a zero-shot scenario. To our knowledge, DeepSearch is also the first deep learning-based peptide identification method capable of conducting zero-shot variable PTM profiling.

\section{Results}
\subsection*{Deep learning-based end-to-end database search}
\begin{figure}[htbp]
    \centering
    \textbf{Database search strategy and the DeepSearch model.}\par\medskip
    \includegraphics[width=0.9\textwidth]{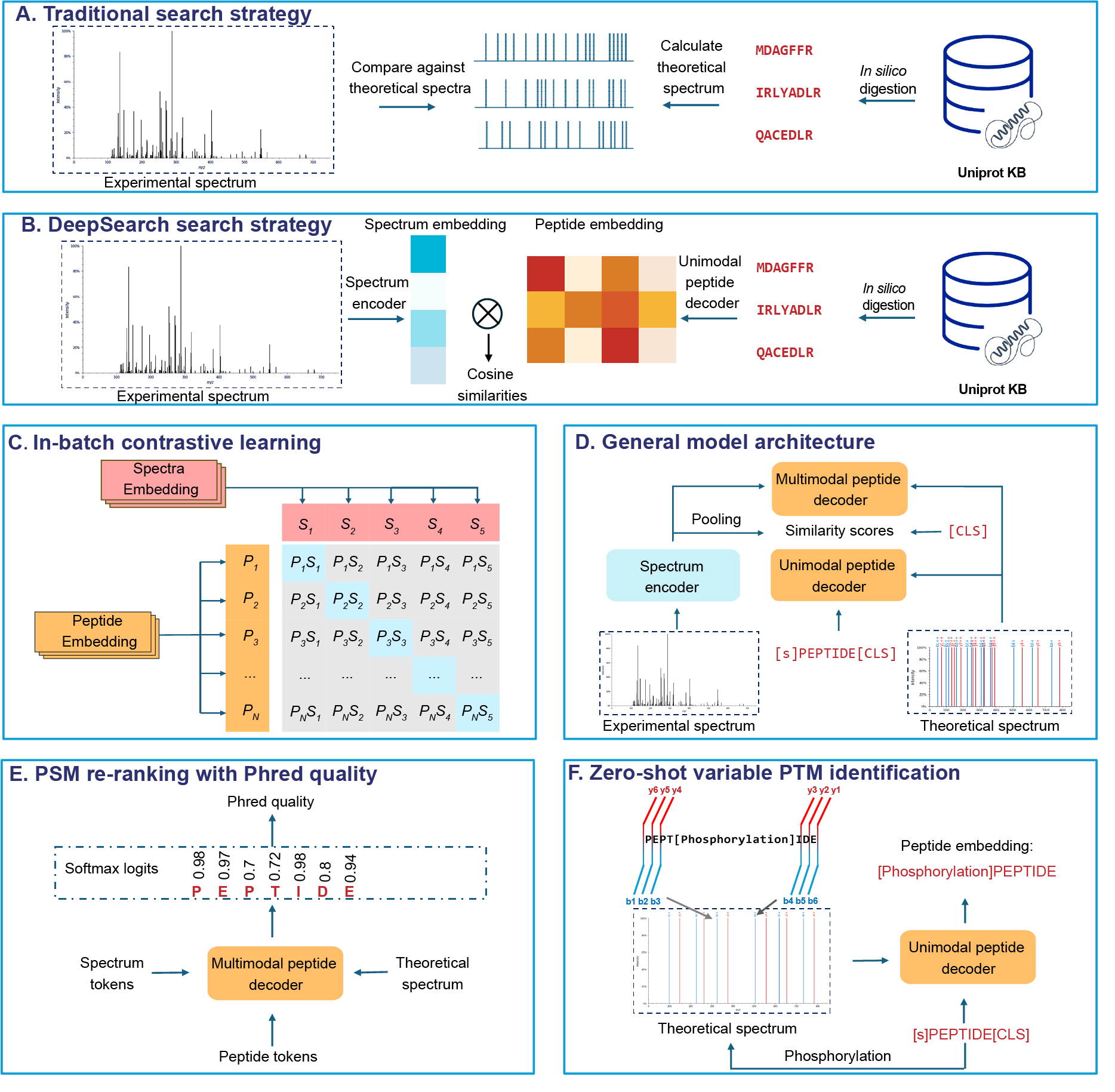}
    \caption{
        A. Conventional database search engines compare experimental MS/MS spectra with theoretical spectra generated from an \textit{in silico} digested peptide database.
        B. DeepSearch performs \textit{in silico} protein digestion and computes an embedding database for peptides.
        Spectrum is encoded into spectrum embedding
        and the cosine similarities between the spectrum embedding and peptide embeddings are computed with matrix multiplication.
        C. DeepSearch uses the in-batch contrastive learning framework without handcrafting negative pairs.
        D. DeepSearch adopts a transformer-based encoder-decoder architecture coupled with a contrastive learning framework.
        E. DeepSearch performs PSM re-ranking with the multimodal peptide decoder using the Phred quality score calculated from softmax probabilities of amino acids in the peptide sequence.
        F. DeepSearch performs zero-shot PTM profiling by shifting the theoretical spectrum with the corresponding PTM mass.
    }\label{fig_main}
\end{figure}

DeepSearch employs a deep learning model to match the experimental MS/MS spectra with peptide sequences, while most conventional database search engines compare experimental MS/MS spectra with theoretical spectra computed from a peptide sequence database (Fig.~\ref{fig_main}A). Starting with an \textit{in silico} digestion of a protein database, DeepSearch encodes the digested peptides and experimental MS/MS spectra into embeddings. Instead of relying on heuristic scoring functions and ion-to-ion matching, DeepSearch uses cosine similarities between corresponding embeddings to score PSMs, which can be efficiently computed through single matrix multiplication (Fig.~\ref{fig_main}B). To address the challenge of annotating closely related negative pairs in PSMs and to mitigate biases from the search engines employed in annotations, we adopted the in-batch contrastive learning framework~\cite{clip,yu2022coca} (Fig.~\ref{fig_main}C). DeepSearch randomly samples a batch of PSMs anchored with peptide mass and utilizes peptide-spectrum pairs, excluding the sampled PSMs, as negative pairs.

We utilized a modified transformer-based encoder-decoder architecture~\cite{transformer, yu2022coca, li2022blip} as shown in Fig. \ref{fig_main}D. The MS/MS spectra were encoded using a spectrum encoder, and spectrum embeddings were derived through a layer of attention pooling~\cite{lee2019set}. Peptide sequences appended with a trainable CLS token and their corresponding theoretical spectra were fed into the unimodal peptide decoder. The embeddings of the CLS token serve as representations of the corresponding peptides. We jointly trained the multimodal peptide decoder with a peptide inference task, which was utilized for PSM re-ranking (Fig. \ref{fig_main}E). We calculated the Phred quality score \cite{phred-sw} for peptide sequences by using the softmax logits corresponding to each amino acid. Additionally, DeepSearch supports PTM profiling without prior training or fine-tuning on PTM enrichment data. Unlike previous methods that encode PTMs as tokens of element compositions \cite{Zeng2022-fv, pdeep2}, we adjusted the theoretical spectrum by the PTM mass to obtain the embedding for peptides with corresponding modifications, as shown in Fig. \ref{fig_main}F. Details are further described in the Methods and Supplementary Note. 

\subsection*{Less biased PSM scoring with DeepSearch}
\begin{figure}[htbp]
    \centering
    \textbf{Search engines reported score distribution by peptide length for \textit{Arabidopsis thaliana} dataset.}\par\medskip
    \includegraphics[width=0.9\textwidth]{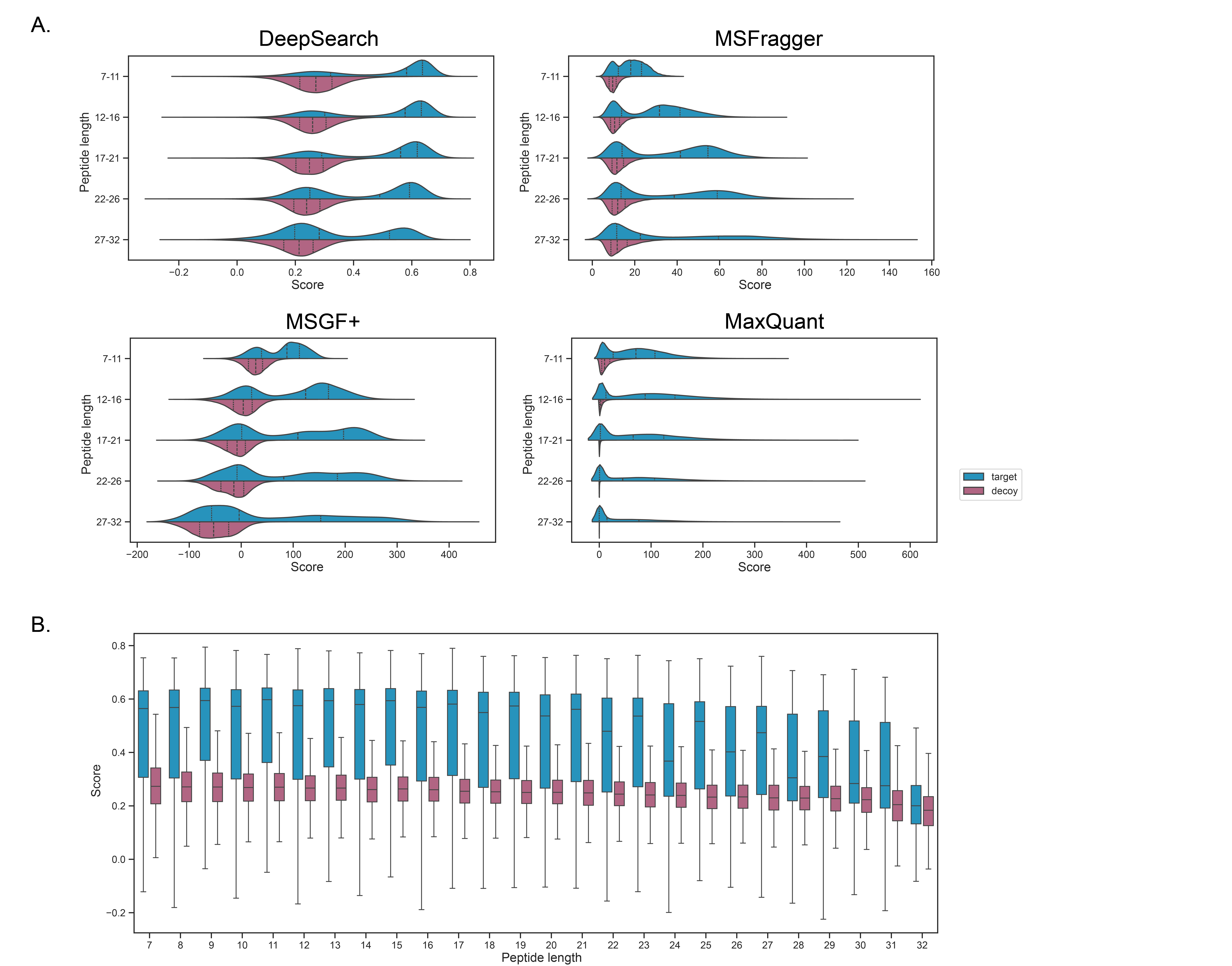}
    \caption{
        A. Search engines reported score distributions for DeepSearch, MSFragger, MSGF+, and MaxQuant. Peptides are grouped into 5 categories based on their length.
        B. DeepSearch reported score distribution by peptide length.
    }\label{fig_score}
\end{figure}

We evaluated the performance of DeepSearch on the proteome-wide higher-energy collisional dissociation (HCD) MS/MS dataset from species with different protein compositions. We compared the results reported by DeepSearch to that of widely adopted search engines MSFragger~\cite{MSFragger-fd}, MSGF+~\cite{msgfplus}, and MaxQuant~\cite{MaxQuant-kc} with a similar search configuration (described in Methods) on \textit{Arabidopsis thaliana}~\cite{Thaliana-cp}, HEK293~\cite{HEK293-ho}, \textit{Caenorhabditis elegans}~\cite{Celegans-wc}, \textit{Escherichia coli}~\cite{ecoli-bb} and HeLa phosphorylation enrichment dataset~\cite{HELA-ba}. We first investigated the scoring function of DeepSearch on peptides of varying lengths on \textit{Arabidopsis thaliana} dataset~\cite{Thaliana-cp}. Traditional database search engines~\cite{MaxQuant-kc, xtandem-ze,comet-vg,MSFragger-fd} with ion-matching scoring functions may exhibit biases towards peptides of different lengths, as longer peptides typically produce more fragmented ions. Conversely, a well-designed scoring function should exhibit reduced sensitivity to peptide compositions and consistently yield similar score distributions for both decoy hits and less confident target hits, assuming the decoys are generated using appropriate strategies~\cite{targetdecoy-qe}. To assess this, we categorized identified peptides into five length-based groups and analyzed their score distributions across all search engines, as depicted in Fig.~\ref{fig_score}. We observed that MSFragger, MSGF+ and MaxQuant tended to assign higher scores to longer peptides, which is foreseeable as longer peptides are more likely to have more matched ions and thus higher scores. Specifically, when the peptide length ranges between 7 and 11 amino acids, the differences between the median scores of target and decoy peptides reported by three other search engines are smaller than those observed in longer peptides. This suggests fewer identifications of shorter peptides if FDR were controlled with scores. Furthermore, we observed that MaxQuant reported significantly fewer decoy hits for longer peptides, indicating a potential bias in its scoring function. In contrast, DeepSearch employs cosine similarities as scoring functions and demonstrates uniform score distributions across all peptide length groups, as shown in Fig~\ref{fig_score}A, B. However, we observed a slight performance drop in peptides longer than 30 amino acids, which can be attributable to the transformer-based model's degradation in handling longer sequences \cite{kazemnejad2023the, li2024functional}. For search engines' score distributions on other benchmarked datasets, see Supplementary Figure 2-4.

\begin{figure}[htbp]
    \centering
    \textbf{Number of PSMs at 1\% FDR on proteome-wide datasets for multiple datasets.}\par\medskip
    \includegraphics[width=0.9\textwidth]{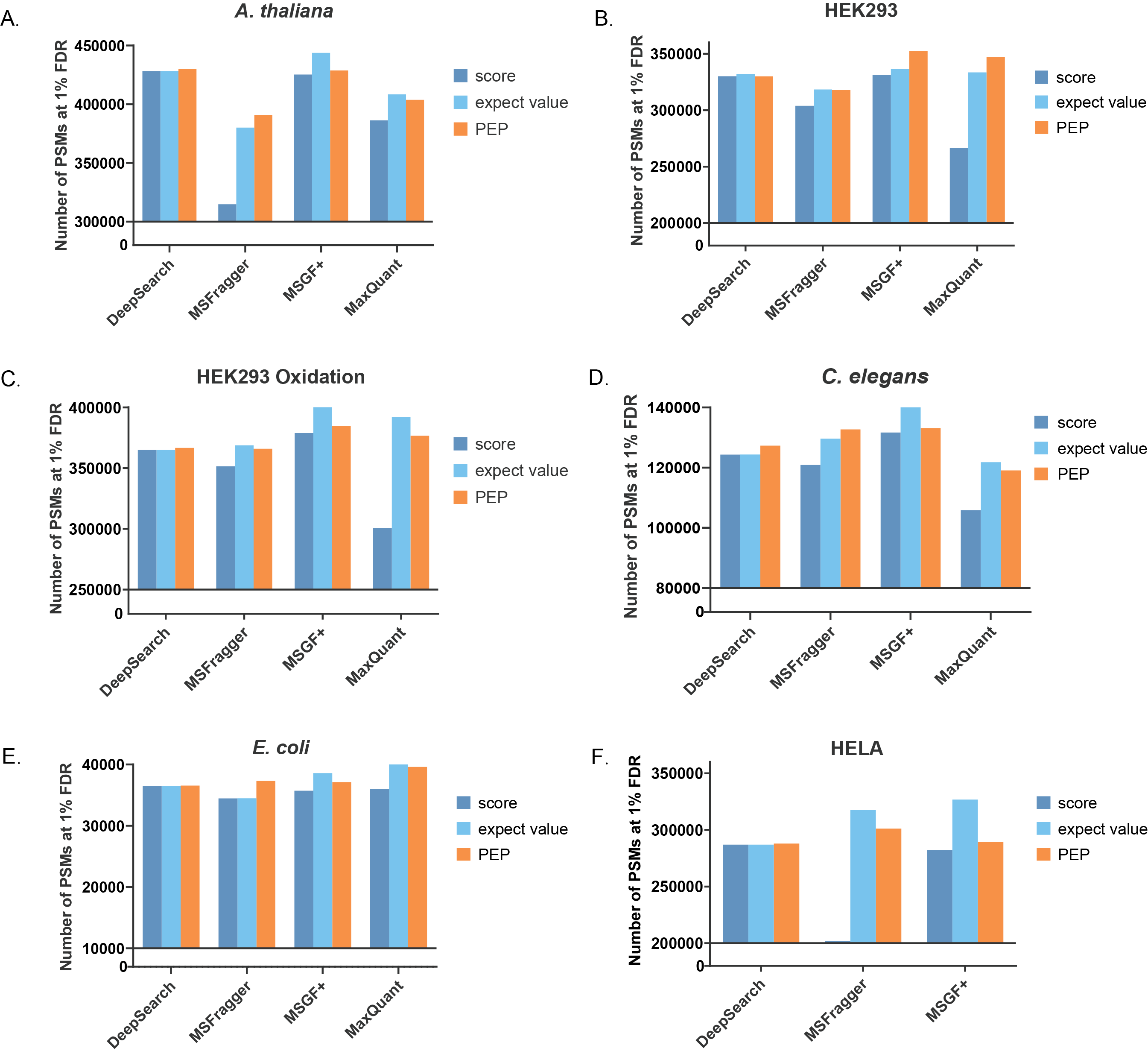}
    \caption{
        PSMs are controlled by raw score, search engine reported expect value, or estimated PEP for
        A. \textit{Arabidopsis thaliana} dataset.
        B. HEK293 dataset.
        C. HEK293 dataset with methionine oxidation as variable PTM.
        D. \textit{Caenorhabditis elegans} dataset.
        E. \textit{Escherichia coli} dataset.
        F. HeLa dataset with methionine oxidation and phosphorylation of serine, threonine, and tyrosine as variable PTMs.
        Different colors represent different search engines, the y-axis represents the number of PSMs with an FDR of 1\%, and MaxQuant failed on the F) HeLa dataset.
    }\label{fig_psm}
    
\end{figure}

To assess the impact of statistical estimations, we benchmarked the number of accepted PSMs at 1\% FDR on the \textit{Arabidopsis thaliana}, HEK293, \textit{Caenorhabditis elegans}, \textit{Escherichia coli} and HeLa phosphorylation enrichment dataset, as shown in Fig.~\ref{fig_psm}. The PSM level FDR control was performed with score, search engine reported expect value (for MaxQuant, this is the reported PEP), or estimated PEP. We observed that DeepSearch constantly reported more PSMs on all datasets than MSFragger and MaxQuant and achieved comparable results with MSGF+ when the FDR was controlled with scores. DeepSearch maintained its consistency on the HEK293 and HeLa datasets when variable PTM search was enabled. 

We noted that these search engines may rely on statistical models to achieve a higher identification rate. We observed significantly fewer reported PSMs at 1\% FDR for MaxQuant when controlled with raw scores across multiple datasets. On the \textit{Arabidopsis thaliana} and HeLa dataset, MSFragger experienced performance degradation without a statistical estimation (Fig.~\ref{fig_psm}A, F). MSGF+ also benefits from its own statistical model for all searches. In contrast, the number of DeepSearch reported PSMs, when controlled with different criteria at 1\% FDR, remains consistent. Such consistency persisted in the peptide and protein levels on all datasets (Table~\ref{table1}, Supplementary Table 2). Further work is necessary to assess the impact of different statistical models, when coupled with target-decoy search strategies, on the quality of the reported matches.

\subsection*{Accurate and robust peptide identification with DeepSearch}
\begin{figure}[htbp]
    \centering
    \textbf{Proteome-wide peptide identifications for the HEK293 dataset.}\par\medskip
    \includegraphics[width=0.9\textwidth]{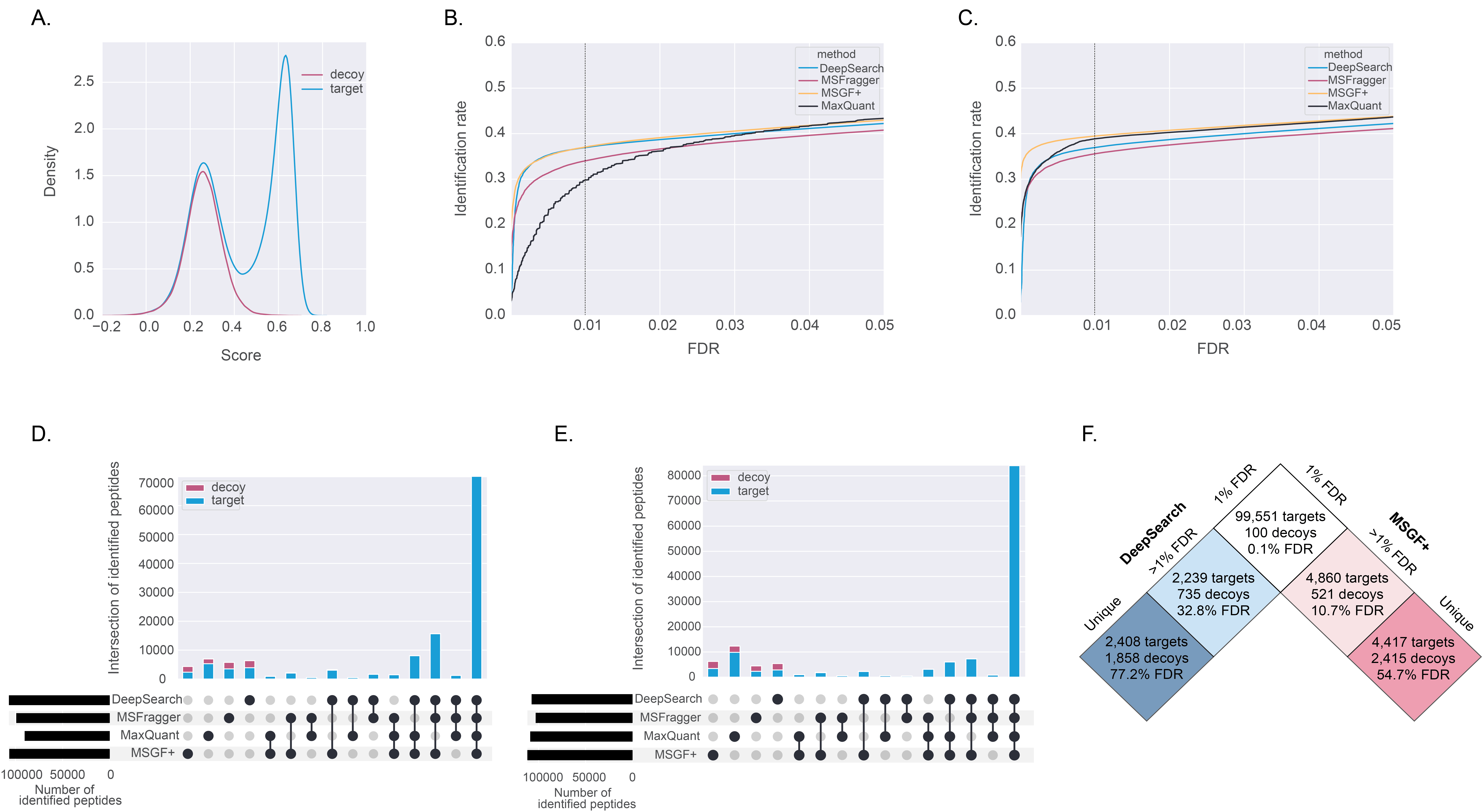}
    \caption{
        A. DeepSearch PSMs' score distribution with target-decoy strategy.
        B. Spectra identification rate against FDR controlled with reported scores for DeepSearch and other search engines.
        C. Spectra identification rate against FDR controlled with reported expect values for DeepSearch and other search engines.
        D. Peptide identification after 1\% PSMs FDR controlled with reported scores for DeepSearch and other search engines.
        E. Peptide identification after 1\% PSMs FDR controlled with reported expect values for DeepSearch and other search engines.
        F. Identified peptides by DeepSearch and MSGF+, divided based on the estimated confidence level. The FDR control is performed with scores for DeepSearch and expect values for MSGF+. Group-specific FDRs are calculated using decoy sequences in each group.
    }\label{figHEK293}
\end{figure}

We conducted a systematic analysis of search engines' reported results on the human proteome-wide HEK293 dataset. Fig.~\ref{figHEK293}A shows DeepSearch reported score distribution, revealing the presence of two distinct clusters corresponding to high and low confidence matches. It is also noticeable that the distributions for low-confident target hits and decoy hits are highly similar despite decoy sequences not being included during training. This similarity is a crucial prerequisite for a well-designed scoring function~\cite{targetdecoy-qe}. We then evaluated the spectra identification rate with FDR controlled using search engines' reported scores or expect values, as shown in Fig.~\ref{figHEK293}B, C. DeepSearch achieved comparable results with MSGF+ and a higher identification rate than MSFragger and MaxQuant with FDR less than 3\% when controlled with scores. When the FDR was performed with expect values, we observed around 6\% and 10\% increase in identified spectra for MSGF+ and MaxQuant correspondingly at 1\% FDR while DeepSearch remains consistent. Fig.~\ref{figHEK293} D, E display the peptide identification for the benchmarked search engines with PSM level FDR controlled at 1\%. When controlled with scores, approximately 89\% of peptides identified by DeepSearch were also reported by at least two other search engines, suggesting DeepSearch's accuracy in identifying PSMs without actual ion matching. This percentage increased to about 92\% when the FDR was controlled using expect values, while DeepSearch identified the same amount of peptides. This further indicated DeepSeach's ability to reliably identify peptides without relying on statistical estimation. Moreover, we observed that MaxQuant relies more on probabilistic estimation to identify some peptides co-found by DeepSearch, MSFragger, and MSGF+. Besides, even with the second-highest spectra identification rate at 1\% FDR when controlled by expect values, MaxQuant still fell short on the number of identified protein groups (Table~\ref{table1}). 

We also examined the intersection of peptides jointly identified by DeepSearch and MSGF+ on the human HEK293 dataset, as the spectra library used for training was constructed with MSGF+. We categorized these peptides based on the confidence level, as shown in Fig.~\ref{figHEK293}F. Peptides accepted by both search engines (99,551 in total) were considered high confidence, with a group FDR of 0.1\%. Peptides identified by both methods but accepted by only one at a 1\% FDR are deemed lower confidence, as suggested by the increased group FDR. Conversely, peptides accepted by one search engine but not identified by the other were considered unique identifications, with the highest group FDR. The 4,860 peptides with lower confidence and 4,417 unique peptides accepted only by MSGF+ suggested that DeepSearch exhibited less bias towards MSGF+ reported results and demonstrated its capability to identify unique peptides.

\begin{figure}[htbp]
    \centering
    \textbf{Proteome-wide peptide identifications for the \textit{Arabidopsis thaliana} dataset.}\par\medskip
    \includegraphics[width=0.9\textwidth]{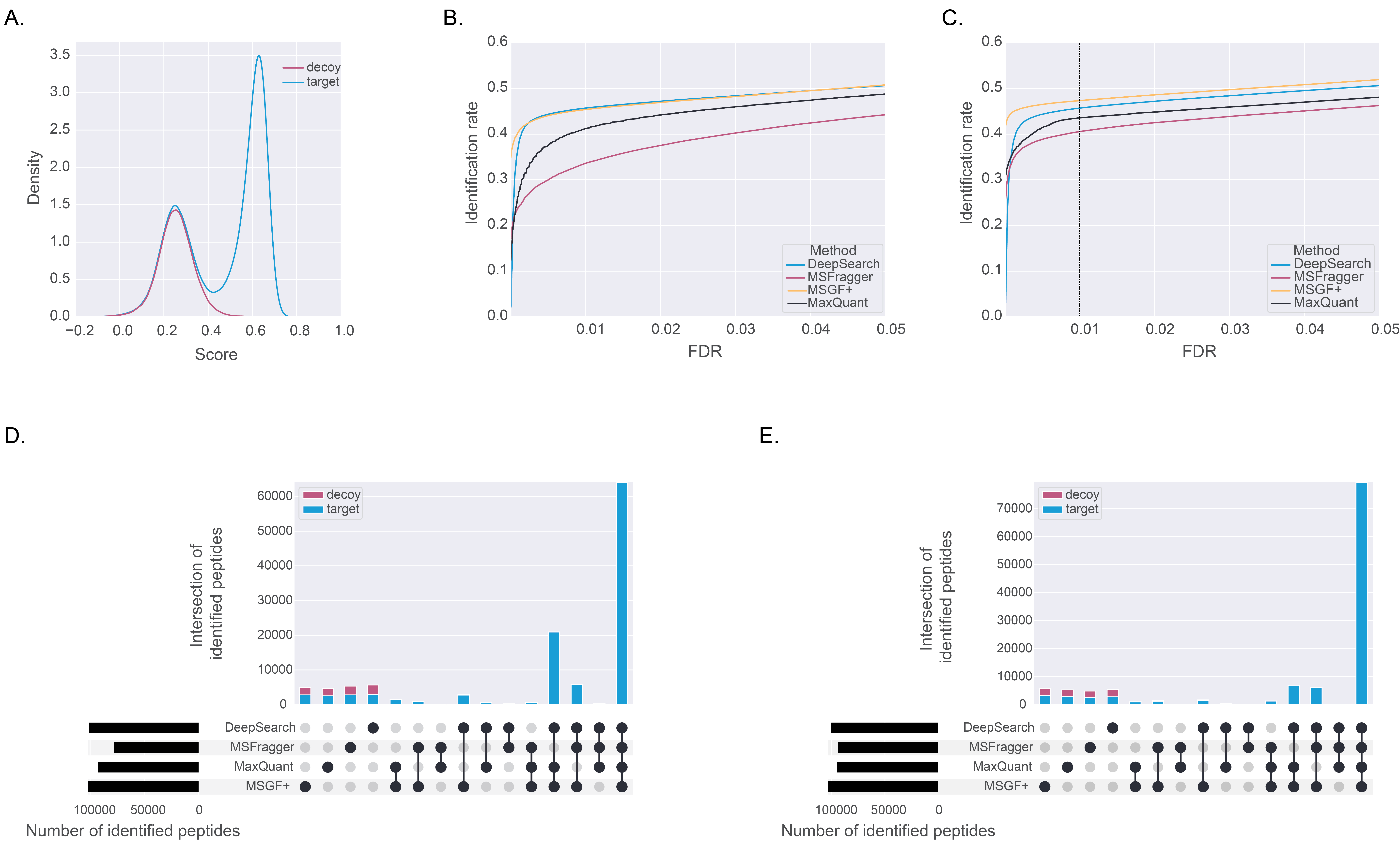}
    \caption{
        A. DeepSearch PSMs' score distribution with target-decoy strategy.
        B. Spectra identification rate against FDR controlled with reported scores for DeepSearch and other search engines.
        C. Spectra identification rate against FDR controlled with reported expect values for DeepSearch and other search engines.
        D. Peptide identification after 1\% PSMs FDR filtering with reported scores for DeepSearch and other search engines.
        E. Peptide identification after 1\% PSMs FDR filtering with reported expect values for DeepSearch and other search engines.
    }\label{fig_tha}
    
\end{figure}

To evaluate the generalization capabilities of DeepSearch on MS/MS spectra from species with a much different protein composition, we further examined reported identifications on the \textit{Arabidopsis thaliana} dataset. Fig.~\ref{fig_tha}A displays the score distribution reported by DeepSearch, and it revealed no significant difference compared to the HEK293 dataset. This suggests DeepSearch's scoring function may generalize effectively across different species. As in the HEK293 dataset, DeepSearch maintained its comparable results regarding the spectra identification rates under both FDR control scenarios (Fig.~\ref{fig_tha}B, C). Fig.~\ref{fig_tha}D, E show the identified peptides reported by all benchmarked search engines at 1\% PSM level FDR. When controlled with PSM scores, around 90.1\% of DeepSearch identified peptides were also reported by at least 2 other search engines. The percentage increases to 92.2\% when controlled with expect values, demonstrating DeepSearch's robustness in accurate peptide identification. Such results imply that DeepSearch is less sensitive to statistical estimations, a pattern also observed in the HEK293 dataset. For benchmarking results of other datasets, see Supplementary Table 2 and Supplementary Figure 5-6.

\begin{table}[htbp]
    \centering
    \caption{Identification results for \textit{Arabidopsis thaliana} and HEK293 dataset at 1\% FDR.}
    \begin{tabular}{ccccccc}

\multicolumn{1}{c|}{Search engine} & \multicolumn{1}{c|}{Peptides} & \multicolumn{1}{c|}{Peptides\textsuperscript{\textdagger}} & \multicolumn{1}{c|}{Peptide\textsuperscript{\textdagger\textdagger}} & \multicolumn{1}{c|}{Proteins} & Proteins\textsuperscript{\textdagger}                   & Proteins\textsuperscript{\textdagger\textdagger} \\ \hline

\multicolumn{7}{l}{Dataset: HEK293}                                                                                                                                                                        \\ \hline
\multicolumn{1}{c|}{DeepSearch}    & \multicolumn{1}{c|}{106891}   & \multicolumn{1}{c|}{106884}   & \multicolumn{1}{c|}{107502}  & \multicolumn{1}{c|}{8467}    & \multicolumn{1}{c|}{8467}     & 8474        \\ \hline
\multicolumn{1}{c|}{MSFragger}     & \multicolumn{1}{c|}{98670}    & \multicolumn{1}{c|}{102595}   & \multicolumn{1}{c|}{104492}  & \multicolumn{1}{c|}{8362}    & \multicolumn{1}{c|}{8447}     & 8344        \\ \hline
\multicolumn{1}{c|}{MSGF+}         & \multicolumn{1}{c|}{106073}   & \multicolumn{1}{c|}{111864}   & \multicolumn{1}{c|}{107215}  & \multicolumn{1}{c|}{8500}    & \multicolumn{1}{c|}{8543}     & 8496        \\ \hline
\multicolumn{1}{c|}{MaxQuant}      & \multicolumn{1}{c|}{89798}    & \multicolumn{1}{c|}{108402}   & \multicolumn{1}{c|}{104503}  & \multicolumn{1}{c|}{7720}    & \multicolumn{1}{c|}{7923}     & 7893        \\ \hline

\multicolumn{7}{l}{Dataset: \textit{A. Thaliana}}                                                                                                                                                          \\ \hline
\multicolumn{1}{c|}{DeepSearch}    & \multicolumn{1}{c|}{100537}   & \multicolumn{1}{c|}{100543}   & \multicolumn{1}{c|}{100975}  & \multicolumn{1}{c|}{10993}     & \multicolumn{1}{c|}{10993} & 11018    \\ \hline
\multicolumn{1}{c|}{MSFragger}     & \multicolumn{1}{c|}{77372}    & \multicolumn{1}{c|}{93756}    & \multicolumn{1}{c|}{99072}   & \multicolumn{1}{c|}{10729}     & \multicolumn{1}{c|}{11069} & 10837    \\ \hline
\multicolumn{1}{c|}{MSGF+}         & \multicolumn{1}{c|}{101894}   & \multicolumn{1}{c|}{103716}   & \multicolumn{1}{c|}{102154}  & \multicolumn{1}{c|}{11037}     & \multicolumn{1}{c|}{11072} & 11041    \\ \hline
\multicolumn{1}{c|}{MaxQuant}      & \multicolumn{1}{c|}{92586}    & \multicolumn{1}{c|}{94647}    & \multicolumn{1}{c|}{94349}   & \multicolumn{1}{c|}{10242}    & \multicolumn{1}{c|}{10339}     & 10296        \\ \hline

\end{tabular}
    Peptide and protein group identifications for the HEK293 and \textit{Arabidopsis thaliana} dataset with DeepSearch and other search engines. PSMs are controlled at 1\% PSM-level FDR. The FDR control is based on search engines' reported scores, expect values (indicated by \textsuperscript{\textdagger}), or estimated PEPs (indicated by \textsuperscript{\textdagger\textdagger}). Proteins are controlled with 1\% protein-level FDR using estimated PEPs for protein groups.
    \label{table1}
\end{table}

\subsection*{Zero-shot variable PTM profiling}
\begin{figure}[htbp]
    \centering
    \textbf{Zero-shot PTM profiling on the phosphorylation enrichment HeLa dataset.}\par\medskip
    \includegraphics[width=0.9\textwidth]{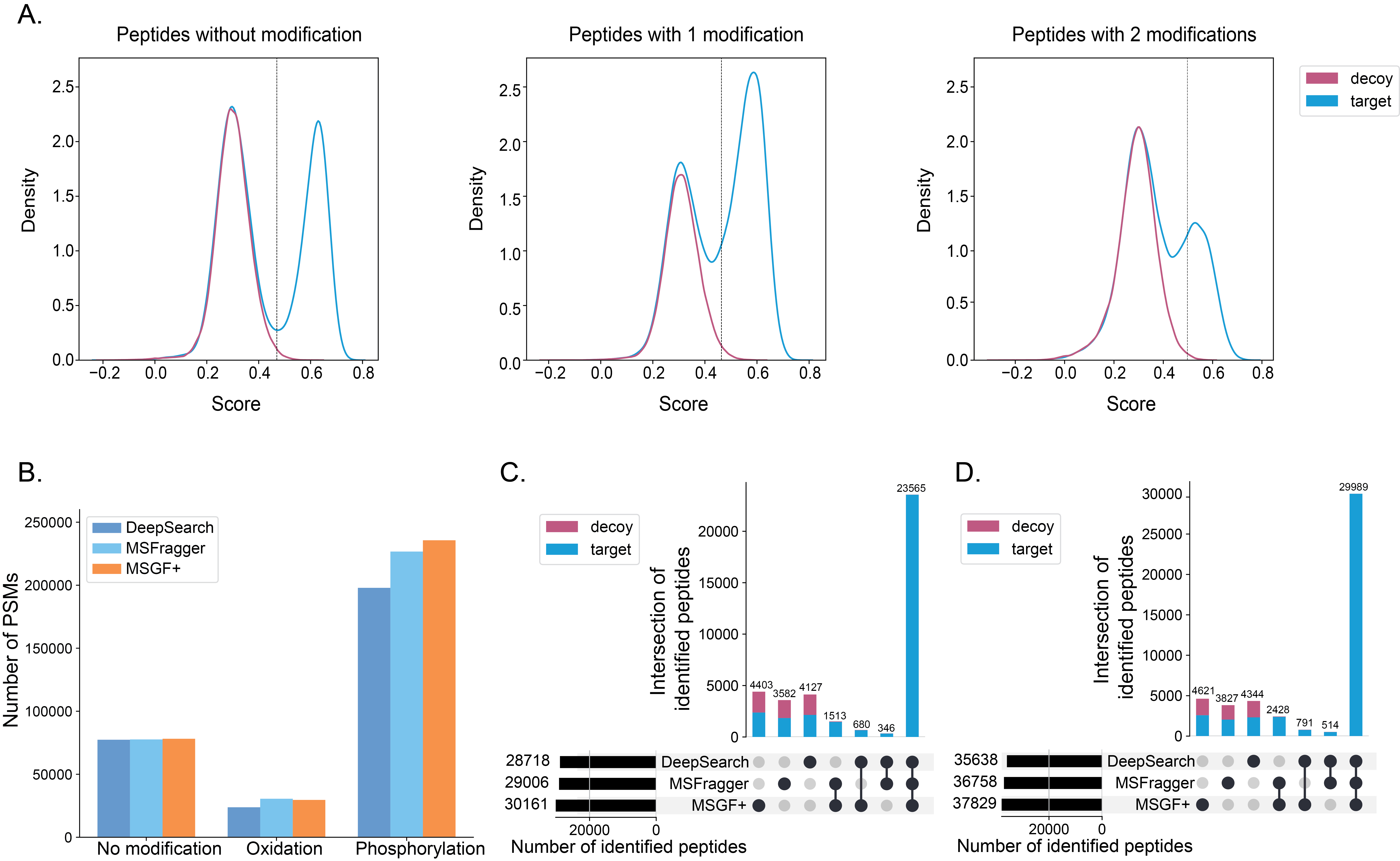}
    \caption{ PTM profiling with methionine oxidation and phosphorylation of serine, threonine, and tyrosine set as variable PTMs. Up to two modifications per peptide are allowed.\\
              A. DeepSearch PSMs' score distribution by the number of modifications, dashed line marks the score at 1\% FDR. 
              B. Number of search engines reported PSMs at 1\% FDR by type of modifications. The FDR is controlled by scores for DeepSearch and expect values for MSFragger and MSGF+. 
              C. Peptide identifications at 1\% PSMs FDR. Peptides were counted in unmodified forms. The FDR is controlled by scores for DeepSearch and expect values for MSFragger and MSGF+. 
              D. Peptide identifications at 1\% PSMs FDR. Peptides were counted with PTM profiles. The FDR is controlled by scores for DeepSearch and expect values for MSFragger and MSGF+. 
    }\label{fig_PTM}
\end{figure}

Deep learning-based peptide identification methods often fall short on variable PTMs identification due to the significant expansion of token space required to encode variable PTMs. Some previous methods only encode methionine oxidation \cite{Mao2023,PointNovo}, which is one of the most common PTMs. Transfer learning has also been applied \cite{Zeng2022-fv, pdeep2} but still requires training on PTM enrichment data. To tackle these problems, DeepSearch introduced a zero-shot variable PTM profiling scheme that effectively generalizes across variable PTMs without enlarging the token space. To evaluate DeepSearch's proficiency in variable PTM profiling, we analyzed its performance using the HeLa phosphorylation enrichment dataset. Fig.~\ref{fig_PTM}A shows DeepSearch's reported score distributions based on the number of modifications in peptides. We observed that with the increased number of modifications, the scores of high-confidence identifications tended to decrease, whereas the score distributions for decoy and low-confidence identifications remained unchanged. This is expected, as performance drops are common in zero-shot learning scenarios. We then examined the PTM profile reported by search engines. DeepSearch achieved comparable results when the FDR was controlled by search engines' scores (Supplementary Figure 7). We also observed fewer PSMs with modifications for DeepSearch compared to those reported by MSFragger and MSGF+ using statistical estimation, highlighting the need for PTM-related probability evaluations (Fig.~\ref{fig_PTM}B). We further examined the accuracy of DeepSearch reported peptides and their PTM profiles (Fig.~\ref{fig_PTM}C, D) at 1\% PSM level FDR. As expected, approximately 82.1\% of peptides and 84.1\% of PTM profiles identified by DeepSearch were also reported by MSFragger and MSGF+, respectively. Nevertheless, there were 1,513 peptides and 2,428 PTM profiles that were jointly reported by MSFragger and MSGF+ but not identified by DeepSearch. For the PTM profiling of the HEK293 dataset with methionine oxidation, see Supplementary Figure 8-9.



\section{Discussion}
In this study, we introduced the first end-to-end deep learning-based database search method, DeepSearch. DeepSearch utilized the modified transformer-based encoder-decoder architectures~\cite{transformer,yu2022coca} to learn a cross-modality embedding space for MS/MS spectra and peptide sequences. To address the challenge of annotating negative pairs of PSMs and biases related to search engine algorithms in the training data, DeepSearch adopted an in-batch contrastive learning framework~\cite{clip, yu2022coca} featuring a mass-anchored sampling scheme. Unlike traditional database search engines which perform ion-to-ion matching, DeepSearch used the cosine similarities between spectra and peptide embeddings to rank the PSMs, which allows for efficient computation through a single matrix multiplication. We evaluated our method on a variety of datasets from species with diverse protein compositions. Though trained only on human spectra library, DeepSearch constantly reported a comparable number of PSMs at 1\% FDR when compared with state-of-the-art database search engines~\cite{MSFragger-fd, MaxQuant-kc, msgfplus} on all datasets. We demonstrated that the majority of peptides identified by DeepSearch were corroborated by other search engines with high confidence. Such results suggest DeepSearch's capability to report peptides accurately and its robustness across species. 

Traditional database search engines depend on heuristic scoring functions, which may be biased towards certain peptide compositions.
These search engines also require statistical estimation~\cite{statical-db, em-len-statical-db} based on the scores to achieve higher identification rates. DeepSearch, on the other hand, employed a data-driven approach to score PSMs. DeepSearch maintained consistent performance, with or without a statistical model, which may be related to its less biased scoring schemes. The impact of statistical estimation, when coupled with the target decoy search strategy, on the quality of PSMs requires further careful examination.

Previous deep learning-based methods in the proteomics field usually fall short on variable PTM profiling since encoding variable PTMs drastically increases the token space~\cite{Zeng2022-fv, pdeep2, GraphNovo}. Moreover, it is impractical to apply transfer learning on PTM enrichment datasets for all common variable PTMs. We demonstrated that DeepSearch was able to report peptides with PTM profiles of high accuracy with phosphorylation and oxidation. Despite this, it requires more examination of DeepSearch's profilings on diverse PTMs. To our knowledge, DeepSearch is the first deep learning-based method capable of zero-shot variable PTM profiling without any prior information other than the mass of the PTM. DeepSearch circumvented the token space limitation by jointly encoding the PTM-shifted theoretical spectrum alongside the unmodified peptide sequence.

\bibliography{sn-bibliography}

\begin{thebibliography}{10}
\providecommand{\url}[1]{{#1}}
\providecommand{\urlprefix}{URL }
\providecommand{\doi}[1]{\url{https://doi.org/#1}}
\bibcommenthead

\bibitem{Nesvizhskii2010-bc}
A.I. Nesvizhskii, A survey of computational methods and error rate estimation procedures for peptide and protein identification in shotgun proteomics.
\newblock J. Proteomics \textbf{73}(11), 2092--2123 (2010)

\bibitem{Aebersold2003-ip}
R.~Aebersold, M.~Mann, Mass spectrometry-based proteomics.
\newblock Nature \textbf{422}(6928), 198--207 (2003)

\bibitem{Eng2011-nc}
J.K. Eng, B.C. Searle, K.R. Clauser, D.L. Tabb, A face in the crowd: recognizing peptides through database search.
\newblock Mol. Cell. Proteomics \textbf{10}(11), R111.009522 (2011)

\bibitem{MaxQuant-kc}
J.~Cox, M.~Mann, {MaxQuant} enables high peptide identification rates, individualized p.p.b.-range mass accuracies and proteome-wide protein quantification.
\newblock Nat. Biotechnol. \textbf{26}(12), 1367--1372 (2008)

\bibitem{MSFragger-fd}
A.T. Kong, F.V. Leprevost, D.M. Avtonomov, D.~Mellacheruvu, A.I. Nesvizhskii, {MSFragger}: ultrafast and comprehensive peptide identification in mass spectrometry-based proteomics.
\newblock Nat. Methods \textbf{14}(5), 513--520 (2017)

\bibitem{msgfplus}
S.~Kim, P.A. Pevzner, {MS-GF+} makes progress towards a universal database search tool for proteomics.
\newblock Nat. Commun. \textbf{5}(1), 5277 (2014)

\bibitem{xtandem-ze}
R.~Craig, R.C. Beavis, A method for reducing the time required to match protein sequences with tandem mass spectra.
\newblock Rapid Commun. Mass Spectrom. \textbf{17}(20), 2310--2316 (2003)

\bibitem{comet-vg}
J.K. Eng, T.A. Jahan, M.R. Hoopmann, Comet: an open-source {MS/MS} sequence database search tool.
\newblock Proteomics \textbf{13}(1), 22--24 (2013)

\bibitem{Liu2020-xf}
K.~Liu, S.~Li, L.~Wang, Y.~Ye, H.~Tang, Full-spectrum prediction of peptides tandem mass spectra using deep neural network.
\newblock Anal. Chem. \textbf{92}(6), 4275--4283 (2020)

\bibitem{statical-db}
D.~Feny{\"o}, R.C. Beavis, A method for assessing the statistical significance of mass spectrometry-based protein identifications using general scoring schemes.
\newblock Anal. Chem. \textbf{75}(4), 768--774 (2003)

\bibitem{em-len-statical-db}
A.~Keller, A.I. Nesvizhskii, E.~Kolker, R.~Aebersold, Empirical statistical model to estimate the accuracy of peptide identifications made by {MS/MS} and database search.
\newblock Anal. Chem. \textbf{74}(20), 5383--5392 (2002)

\bibitem{DeepNovo-uy}
N.H. Tran, X.~Zhang, L.~Xin, B.~Shan, M.~Li, De novo peptide sequencing by deep learning.
\newblock Proc. Natl. Acad. Sci. U. S. A. \textbf{114}(31), 8247--8252 (2017)

\bibitem{PointNovo}
R.~Qiao, N.H. Tran, L.~Xin, X.~Chen, M.~Li, B.~Shan, A.~Ghodsi, Computationally instrument-resolution-independent de novo peptide sequencing for high-resolution devices.
\newblock Nature Machine Intelligence \textbf{3}(5), 420--425 (2021).
\newblock \doi{10.1038/s42256-021-00304-3}.
\newblock \urlprefix\url{https://doi.org/10.1038/s42256-021-00304-3}

\bibitem{pmlr-casanovo}
M.~Yilmaz, W.~Fondrie, W.~Bittremieux, S.~Oh, W.S. Noble, \emph{De novo mass spectrometry peptide sequencing with a transformer model}, in \emph{Proceedings of the 39th International Conference on Machine Learning}, \emph{Proceedings of Machine Learning Research}, vol. 162, ed. by K.~Chaudhuri, S.~Jegelka, L.~Song, C.~Szepesvari, G.~Niu, S.~Sabato (PMLR, 2022), pp. 25514--25522.
\newblock \urlprefix\url{https://proceedings.mlr.press/v162/yilmaz22a.html}

\bibitem{Mao2023}
Z.~Mao, R.~Zhang, L.~Xin, M.~Li, Mitigating the missing-fragmentation problem in de novo peptide sequencing with a two-stage graph-based deep learning model.
\newblock Nature Machine Intelligence \textbf{5}(11), 1250--1260 (2023).
\newblock \doi{10.1038/s42256-023-00738-x}.
\newblock \urlprefix\url{https://doi.org/10.1038/s42256-023-00738-x}

\bibitem{qi2017pointnet}
C.R. Qi, H.~Su, K.~Mo, L.J. Guibas, Pointnet: Deep learning on point sets for 3d classification and segmentation  (2017).
\newblock {\href{https://arxiv.org/abs/1612.00593}{{arXiv:1612.00593}}} {[cs.CV]}

\bibitem{prosit_trans}
M.~Ekvall, P.~Truong, W.~Gabriel, M.~Wilhelm, L.~Käll, Prosit transformer: A transformer for prediction of ms2 spectrum intensities.
\newblock Journal of Proteome Research \textbf{21}(5), 1359--1364 (2022).
\newblock \doi{10.1021/acs.jproteome.1c00870}.
\newblock \urlprefix\url{https://doi.org/10.1021/acs.jproteome.1c00870}.
\newblock PMID: 35413196.
\newblock {\href{https://arxiv.org/abs/https://doi.org/10.1021/acs.jproteome.1c00870}{{https://doi.org/10.1021/acs.jproteome.1c00870}}}

\bibitem{Ramazi2021-fy}
S.~Ramazi, J.~Zahiri, Posttranslational modifications in proteins: resources, tools and prediction methods.
\newblock Database (Oxford) \textbf{2021} (2021)

\bibitem{clip}
A.~Radford, J.W. Kim, C.~Hallacy, A.~Ramesh, G.~Goh, S.~Agarwal, G.~Sastry, A.~Askell, P.~Mishkin, J.~Clark, G.~Krueger, I.~Sutskever, \emph{Learning Transferable Visual Models From Natural Language Supervision}, in \emph{Proceedings of the 38th International Conference on Machine Learning}, \emph{Proceedings of Machine Learning Research}, vol. 139, ed. by M.~Meila, T.~Zhang (PMLR, 2021), pp. 8748--8763.
\newblock \urlprefix\url{https://proceedings.mlr.press/v139/radford21a.html}

\bibitem{li2022blip}
J.~Li, D.~Li, C.~Xiong, S.~Hoi, \emph{Blip: Bootstrapping language-image pre-training for unified vision-language understanding and generation}, in \emph{International conference on machine learning} (PMLR, 2022), pp. 12888--12900

\bibitem{yu2022coca}
J.~Yu, Z.~Wang, V.~Vasudevan, L.~Yeung, M.~Seyedhosseini, Y.~Wu, Coca: Contrastive captioners are image-text foundation models.
\newblock Transactions on Machine Learning Research  (2022).
\newblock \urlprefix\url{https://openreview.net/forum?id=Ee277P3AYC}

\bibitem{onetransformer}
F.~Bao, S.~Nie, K.~Xue, C.~Li, S.~Pu, Y.~Wang, G.~Yue, Y.~Cao, H.~Su, J.~Zhu, \emph{One transformer fits all distributions in multi-modal diffusion at scale}, in \emph{Proceedings of the 40th International Conference on Machine Learning} (JMLR.org, 2023), ICML'23

\bibitem{ALIGN}
C.~Jia, Y.~Yang, Y.~Xia, Y.T. Chen, Z.~Parekh, H.~Pham, Q.~Le, Y.H. Sung, Z.~Li, T.~Duerig, \emph{Scaling Up Visual and Vision-Language Representation Learning With Noisy Text Supervision}, in \emph{Proceedings of the 38th International Conference on Machine Learning}, \emph{Proceedings of Machine Learning Research}, vol. 139, ed. by M.~Meila, T.~Zhang (PMLR, 2021), pp. 4904--4916.
\newblock \urlprefix\url{https://proceedings.mlr.press/v139/jia21b.html}

\bibitem{wang2022simvlm}
Z.~Wang, J.~Yu, A.W. Yu, Z.~Dai, Y.~Tsvetkov, Y.~Cao, \emph{Sim{VLM}: Simple Visual Language Model Pretraining with Weak Supervision}, in \emph{International Conference on Learning Representations} (2022).
\newblock \urlprefix\url{https://openreview.net/forum?id=GUrhfTuf\_3}

\bibitem{MassIVE}
M.~Wang, J.~Wang, J.~Carver, B.S. Pullman, S.W. Cha, N.~Bandeira, Assembling the community-scale discoverable human proteome.
\newblock Cell Syst. \textbf{7}(4), 412--421.e5 (2018)

\bibitem{transformer}
A.~Vaswani, N.~Shazeer, N.~Parmar, J.~Uszkoreit, L.~Jones, A.N. Gomez, L.u. Kaiser, I.~Polosukhin, \emph{Attention is All you Need}, in \emph{Advances in Neural Information Processing Systems}, vol.~30, ed. by I.~Guyon, U.V. Luxburg, S.~Bengio, H.~Wallach, R.~Fergus, S.~Vishwanathan, R.~Garnett (Curran Associates, Inc., 2017)

\bibitem{lee2019set}
J.~Lee, Y.~Lee, J.~Kim, A.~Kosiorek, S.~Choi, Y.W. Teh, \emph{Set Transformer: A Framework for Attention-based Permutation-Invariant Neural Networks}, in \emph{Proceedings of the 36th International Conference on Machine Learning}, \emph{Proceedings of Machine Learning Research}, vol.~97, ed. by K.~Chaudhuri, R.~Salakhutdinov (PMLR, 2019), pp. 3744--3753.
\newblock \urlprefix\url{https://proceedings.mlr.press/v97/lee19d.html}

\bibitem{phred-sw}
B.~Ewing, L.~Hillier, M.C. Wendl, P.~Green, Base-calling of automated sequencer traces using phred. i. accuracy assessment.
\newblock Genome Res. \textbf{8}(3), 175--185 (1998)

\bibitem{Zeng2022-fv}
W.F. Zeng, X.X. Zhou, S.~Willems, C.~Ammar, M.~Wahle, I.~Bludau, E.~Voytik, M.T. Strauss, M.~Mann, {AlphaPeptDeep}: a modular deep learning framework to predict peptide properties for proteomics.
\newblock Nat. Commun. \textbf{13}(1), 7238 (2022)

\bibitem{pdeep2}
W.F. Zeng, X.X. Zhou, W.J. Zhou, H.~Chi, J.~Zhan, S.M. He, {MS/MS} spectrum prediction for modified peptides using pdeep2 trained by transfer learning.
\newblock Anal. Chem. \textbf{91}(15), 9724--9731 (2019)

\bibitem{Thaliana-cp}
H.~Zhang, P.~Liu, T.~Guo, H.~Zhao, D.~Bensaddek, R.~Aebersold, L.~Xiong, Arabidopsis proteome and the mass spectral assay library.
\newblock Sci. Data \textbf{6}(1), 278 (2019)

\bibitem{HEK293-ho}
J.M. Chick, D.~Kolippakkam, D.P. Nusinow, B.~Zhai, R.~Rad, E.L. Huttlin, S.P. Gygi, A mass-tolerant database search identifies a large proportion of unassigned spectra in shotgun proteomics as modified peptides.
\newblock Nat. Biotechnol. \textbf{33}(7), 743--749 (2015)

\bibitem{Celegans-wc}
D.M. Walther, P.~Kasturi, M.~Zheng, S.~Pinkert, G.~Vecchi, P.~Ciryam, R.I. Morimoto, C.M. Dobson, M.~Vendruscolo, M.~Mann, F.U. Hartl, Widespread proteome remodeling and aggregation in aging c. elegans.
\newblock Cell \textbf{161}(4), 919--932 (2015)

\bibitem{ecoli-bb}
A.~Schmidt, K.~Kochanowski, S.~Vedelaar, E.~Ahrn{\'e}, B.~Volkmer, L.~Callipo, K.~Knoops, M.~Bauer, R.~Aebersold, M.~Heinemann, The quantitative and condition-dependent escherichia coli proteome.
\newblock Nat. Biotechnol. \textbf{34}(1), 104--110 (2016)

\bibitem{HELA-ba}
K.~Sharma, R.C.J. D'Souza, S.~Tyanova, C.~Schaab, J.R. Wi{\'s}niewski, J.~Cox, M.~Mann, Ultradeep human phosphoproteome reveals a distinct regulatory nature of tyr and {Ser/Thr-based} signaling.
\newblock Cell Rep. \textbf{8}(5), 1583--1594 (2014)

\bibitem{targetdecoy-qe}
J.E. Elias, S.P. Gygi, Target-decoy search strategy for increased confidence in large-scale protein identifications by mass spectrometry.
\newblock Nat. Methods \textbf{4}(3), 207--214 (2007)

\bibitem{kazemnejad2023the}
A.~Kazemnejad, I.~Padhi, K.~Natesan, P.~Das, S.~Reddy, \emph{The Impact of Positional Encoding on Length Generalization in Transformers}, in \emph{Thirty-seventh Conference on Neural Information Processing Systems} (2023).
\newblock \urlprefix\url{https://openreview.net/forum?id=Drrl2gcjzl}

\bibitem{li2024functional}
S.~Li, C.~You, G.~Guruganesh, J.~Ainslie, S.~Ontanon, M.~Zaheer, S.~Sanghai, Y.~Yang, S.~Kumar, S.~Bhojanapalli, \emph{Functional Interpolation for Relative Positions improves Long Context Transformers}, in \emph{The Twelfth International Conference on Learning Representations} (2024).
\newblock \urlprefix\url{https://openreview.net/forum?id=rR03qFesqk}

\bibitem{GraphNovo}
Z.~Mao, R.~Zhang, L.~Xin, M.~Li, Mitigating the missing-fragmentation problem in de novo peptide sequencing with a two-stage graph-based deep learning model.
\newblock Nature Machine Intelligence \textbf{5}(11), 1250--1260 (2023).
\newblock \doi{10.1038/s42256-023-00738-x}.
\newblock \urlprefix\url{https://doi.org/10.1038/s42256-023-00738-x}

\end{thebibliography}

\end{document}